\documentclass{article}
\usepackage{amssymb}
\usepackage{amsmath}

\usepackage{amsfonts}
\usepackage{epsfig}

\newtheorem{theorem}{Theorem}
\newtheorem{acknowledgement}[theorem]{Acknowledgement}

\begin{document}

\title{A non-singular black hole model as a possible end-product of
gravitational collapse.}
\author{\thanks{%
\ \ NAS/NRC Senior Research Associate;~~e-mail: mrmsps@rit.edu}\ Manasse R. Mbonye$^{1,2}$ and Demosthenes
Kazanas$^{1}$ \\
\ \ \ \ \\
$^{1}$\textit{NASA/Goddard Space Flight Center}\\
\textit{Mail Code 663,}\\
\textit{Greenbelt, MD 20785}\\
\ \ \ \ \ \ \\
$^{2}$\textit{Department of Physics, }\\
\textit{Rochester Institute of Technology, }\\
\textit{84 Lomb Drive, Rochester, NY 14623.} }
\maketitle

\begin{abstract}
In this paper we present a non-singular black hole model as a possible
end-product of gravitational collapse. The depicted spacetime which is type $%
\left[ II,\left( II\right) \right] $, by Petrov classification, is an exact
solution of the Einstein equations with and contains two horizons. The
equation of state $p_{r}\left( \rho \right) $ in the radial direction, is a
well-behaved function of the density $\rho \left( r\right) $ and smoothly
reproduces vacuum-like behavior near $r=0$ while tending to a polytrope at
larger $r$, low $\rho $ values. The final equilibrium configuration
comprises a de Sitter-like inner core surrounded by a family of 2-surfaces $%
\Sigma $ of matter fields with variable equation of state. The fields are
all concentrated in the vicinity of the radial center $r=0$. The solution
depicts a spacetime that is asymptotically Schwarzschild at large $r$, while
it becomes de Sitter-like as $r\rightarrow 0$. Possible physical
interpretations of the macro-state of the black hole interior in the model
are offered. We find that the possible state admits two equally viable
interpretations, namely either a quintessential intermediary region or a
phase transition in which a two-fluid system is in both dynamic and
thermodynamic equilibrium. We estimate the ratio of pure matter present to
the total energy and in both cases find it to be virtually the same, being $%
\sim 0.83$. Finally, the well-behaved dependence of the density and pressure
on the radial coordinate provides some insight on dealing with the
information loss paradox.
\end{abstract}


\textbf{Pacs: \ \ \ \ \ \ \ \ \ }04.70.Bw, 04.20.Jb

\

\section{Introduction}

It is generally accepted that the formation of a black hole proceeds with
the formation of a marginally trapped surface (the apparent horizon) into
which matter is collapsing and one that encloses a region of space interior
to which the light cones are tipped [1] so much so that even outgoing null
geodesics proceed toward ever decreasing values of the radial coordinate.
The famous singularity theorems [2][3][4], then posit that under certain
assumptions on energy conditions of matter, the eventual state of the
collapse is an infinite density singularity surrounded by an event horizon
(defined by the outermost trapped surface). It is, however, a fact that the
existence of such a singularity is neither easy to verify nor refute
observationally because of the limits set by the black hole event horizon.
In the absence of such observational evidence, the compelling logic in the
arguments leading to the singularity theorems has generally shaped the
widespread conviction that indeed a black hole must contain a physical
singularity.

As is known, however, the picture of a singular black hole opens up several
puzzles. For example, the existence of spacetime singularities results in
irretrievably total information loss [5]. Moreover, the infinite tidal
forces that result from such a singularity formation lead to a breakdown in
the descriptive power of General Relativity [6]. As a result it is generally
believed that the perceived final collapse state implied by the singularity
theorems could be but a manifestation of the incompleteness of classical
gravity and that a correct (quantum) theory of gravity will dispense with
this singular final state and admit a state of finite (albeit very high)
curvature. An altogether different view is that the singularities implied by
the aforementioned theorems simply reflect our lack of knowledge of the
properties of matter under extreme conditions. Thus it has been suggested
that the singularity theorems could be circumvented if (some of) the
conditions imposed on the properties of collapsing matter were relaxed.
Indeed modern frameworks (like string theory) attempting to formulate a
quantum theory of gravity seek the existence of a fundamental length scale
and hence a singularity free spacetime. There are also philosophical
questions as to whether spacetime can indeed accommodate singularities of
any kind, be they from gravitational collapse or vacuum-induced [7].

As a result of these and other considerations a renewed interest has grown
lately with regard to spacetime singularities and their existence [8].
Indeed the concept of non-singular collapse dates back to Sakharov's
consideration of the equation of state $p=-\rho$ for a superdense fluid and
Gliner's view [9] that such a fluid could be the final state of
gravitational collapse. Such ideas are now attracting a growing amount of
attention. For example, Ellis [10] has argued that in the case of a closed
universe a trapped surface does not necessarily lead to a singularity.
Recently the present authors have investigated an analogous argument with
regard to black holes in [11]. It was pointed out in [11] that under
physically acceptable energy conditions the expansion of the outgoing mode
which assumes increasingly negative values in a trapped region can
eventually turn around to (even) assume a positive signature inside a black
hole. Such behavior suggests the existence some interior region, surrounding
the core, which is not trapped. In turn, this feature implies the absence of
a singularity in such a spacetime. In recent years, several non-singular
black hole solutions have been found. Dymnikova, for example, [12] has
constructed an exact solution of the Einstein field equations for a
non-singular black hole containing at the core a fluid de-Sitter-like fluid
with anisotropic pressure. This solution is Petrov Type [(II),(II)] by
classification\footnote{%
We shall present a solution with the same asymptotic behavior as in [12]
which however is different in both classification (it is Petrov Type
[II,(II)] and physical interpretation.}. Another line of investigation in
this area, initiated by Markov [13], suggests a limiting curvature approach.
This idea has been explored further\ by several authors, see for example
[14][15][16][17][18][19].

A common feature in all these treatments is that the geometry of the
spacetime in question is Schwarschild at large $r$ as expected by Birkoff's
theorem, and de Sitter-like at small $r$\ values as implied by the $p=-\rho $
equation of state assumed to prevail near $r=0$. These approaches can be put
in two broad classes. (a) Those for which the transition to this
\textquotedblleft exotic" state of matter can \ be placed well inside the
Schwarzschild horizon [12] and (b) those which replace the entire volume of
the Schwarschild metric interior to $r=2M$ by a substance with the $p=-\rho $
equation of state [6], dispensing in this way with the presence of the
horizon altogether and the \textquotedblleft information paradox" problems
it engenders.

The problem of transition from the usual matter equation of state to that
appropriate for these non-singular solutions is usually not addressed at
all. This problem is particularly acute for models of the second class
discussed above, since the density at horizon formation scales like $%
\rho\simeq 10^{16}\,(M/M_{\odot})^{-2}$ g cm$^{-3}$ and for objects of mass $%
M\geq 10^{8}\,M_{\odot})$, is not greater than that of water, which is known
to have an equation of state very different from $p=-\rho$. Moreover, issues
to do with direct matching of an external Schwarschild vacuum to an interior
de Sitter have previously been discussed [20] based on junction conditions
[21]. In fact, in terms of gravitational collapse the associated
difficulties have been used to suggest modifications in some treatments
[22]. The present work is motivated by these questions and is undertaken as
a first effort to provide some answers by constructing models using an
explicit equation of state with the desired properties. It would appear that
the junction constraints [21] when applied to the end-product of
non-singular gravitational collapse imply the matter fields across the
Schwarschild/de Sitter boundary should have a radially dependent equation of
state of the form $1\lesssim w(r)\lesssim -1$\ that smoothly changes the
matter-energy from a stiff fluid \ to a cosmological constant.

The model sketched in this paper describes the geometry of a body that
passed during its collapse through a stiff fluid state to settle into the
final state of a non-singular black hole. The solution allows for a matter
fluid region (with a family of equations of state $0\leq w\left( r\right)
\leq 1$ which envelopes an intermediate region with a family of equations of
state $-1<w\left( r\right) <0$ which in turn envelopes a de Sitter like
region with $w=-1$ at the center. Our model differs both from the
traditional singular black hole solutions which introduce matter at the
singularity and from the relatively new non-singular solutions which usually
introduced only a de sitter-like spacetime inside the black hole. The main
feature of our model is that it

(1) introduces gravitating matter inside a non-singular black hole and

(2) offers a reasonable explanation of how part of this matter can evolve
towards a de Sitter-like vacuum to provide the radial tension or negative
pressure that supports the remaining matter fields against the formation of
a singularity.

The rest of the paper is organized as follows. In section 2 we summarize the
equations to be solved. We also construct the working equation of state and
discuss some of its desirable features. In section 3 we solve the Einstein
Field Equations for the desired geometry and obtain an exact solution. In
section 4 we highlight the physical implications of the model. Section 5
concludes the discussion.

\section{Problem formulation}

\subsection{The field equations}

It is assumed, for simplicity, that the collapsed object can be reasonably
described by a spherical, static geometry. The desired line element then
takes the general form 
\begin{equation}
ds^{2}=-e^{\nu\left( r\right) }dt^{2}+e^{\lambda\left( r\right)
}dr^{2}+r^{2}\left( d\theta^{2}+\sin^{2}\theta d\varphi^{2}\right) ,  
\tag{2.1}
\end{equation}
where $\lambda\left( r\right) $ and $\nu\left( r\right) $ are to be
determined from the Einstein Field Equations 
\begin{equation}
G_{\mu\nu}=-8\pi GT_{\mu\nu}.   \tag{2.2}
\end{equation}

\bigskip Under the assumed spacetime symmetry, Eqs.2 reduce to%
\begin{equation}
e^{-\lambda}\left( \frac{\lambda^{\prime}}{r}-\frac{1}{r^{2}}\right) +~\frac{%
1}{r^{2}}=8\pi G\rho\left( r\right) ,   \tag{2.3}
\end{equation}

\begin{equation}
e^{-\lambda}\left( \frac{\nu^{\prime}}{r}+\frac{1}{r^{2}}\right) -~\frac {1}{%
r^{2}}=8\pi Gp_{r}\left( r\right) ,   \tag{2.4}
\end{equation}

and 
\begin{equation}
e^{-\lambda}\left( \frac{\nu^{\prime\prime}}{2}-\frac{\lambda^{\prime}\nu^{%
\prime}}{4}+\frac{\nu^{\prime2}}{4}+\frac{\nu^{\prime}-\lambda^{\prime}}{2r}%
\right) =8\pi p_{\perp}\left( r\right) .   \tag{2.5}
\end{equation}
Here $\rho\left( r\right) =$\ $T_{0}^{0}$ is the energy density, $%
p_{r}=T_{1}^{1}$ is the radial pressure and $\left(
p_{\theta}=T_{2}^{2}\right) =\left( p_{\varphi}=T_{3}^{3}\right) =p_{\perp}$
is the tangential pressure. In this treatment the fluid is anisotropic with $%
T_{2}^{2}=T_{3}^{3}\neq T_{1}^{1}$ and $T_{1}^{1}\neq T_{0}^{0}\neq
T_{2}^{2} $. The spacetime is thus Petrov Type [II, (II)], where ( ) implies
a degeneracy in the eigenvalues of the\ Weyl tensor.

\subsection{Equation of state}

In the remaining part of this section we construct the working equation of
state $p_{r}\left( \rho\right) $ in the radial direction which reproduces
the characteristics of the collapsed body as highlighted in the previous
section. We justify the choice by pointing out its desired features. On the
other hand, the tangential pressure equation of state $p_{\perp}\left(
\rho\right) $ will be derived later from the Einstein Field Equations (Eq.
2.5) using the radial pressure function $p_{r}\left( \rho\right) $ to be
constructed.

The main desired features are that (1) the equation of $p_{r}\left(
\rho\right) $ change from the one associated with the usual matter, at low
densities, to that associated with the de Sitter geometry at sufficiently
high densities near the center of the configuration and; (2) that this
change be well-behaved. To this end we consider an equation of state that
takes the general form\footnote{%
This choice may not be unique.} where $m$ and $1/n$ are, yet to be
determined, positive real numbers and $\alpha$ is a free parameter to be
constrained. This equation implies presence of a%
\begin{equation}
p_{r}\left( \rho\right) =\left[ \alpha-\left( \alpha+1\right) \left( \frac{%
\rho}{\rho_{\max}}\right) ^{m}\right] \left( \frac{\rho}{\rho_{\max}}\right)
^{1/n}\rho,   \tag{2.6}
\end{equation}
maximum limiting density $\rho_{\max}$ concentrated in a region of size $%
r_{0}=\sqrt{\frac{1}{G\rho_{\max}}}$, the core region of the configuration.
One notes that the main desirable features are readily apparent in the
functional form of Eq. 2.6. At low densities, $\left( \frac{\rho}{\rho_{\max
}}\right) ^{m}<\frac{\alpha}{\alpha+1}$, the equation of state reduces to
that of a polytrope of index $n$ ($p_{r}\propto\rho^{1+1/n}$) while for $%
\left( \frac{\rho}{\rho_{\max}}\right) ^{m}>\frac{\alpha}{\alpha+1}$ the
pressure decreases to eventually approach that of the vacuum $%
p_{r}=-\rho_{\max}$ as $\rho\rightarrow\rho_{\max}$. We look for the
simplest form of Eq. 2.6 by judiciously constraining both the indices $m$
and $n$ and the parameter $\alpha$.

Since Eq. 6 already satisfies the general features desired by the model, its
specific form, based on an appropriate choice of the indices $m$ and $n$ and
the parameter $\alpha$ is made through the demand that it must satisfy the
following basic conditions.

(1) It must have no pathologies, i.e.

\ \ \ \ (i) the sound speed $\frac{dp}{d\rho}$ can not be a maximum at $%
\rho=0$

\ \ \ \ \ (ii) in order to rule out superluminal behavior, $\frac{dp}{d\rho }%
\ngtr1$, the maximum sound speed, i.e. the value of $\frac{dp}{d\rho}$ at
the point where $\frac{d^{2}p}{d\rho^{2}}=0$ must be given by $\frac{dp}{%
d\rho }\mid_{\rho_{stiff}}=c_{s}^{2}=c=1$.

(2) It must satisfy minimum acceptable energy conditions, namely the Weak
Energy Condition, $\rho\geq0$, $\rho+p_{r}\geq0$ and the Dominant energy
Condition, $\rho\geq0$, $p_{r}\in\left[ -\rho,^{+}\rho\right] $.

The $\left[ m=1,1/n=0\right] $ and the $\left[ m=2,1/n=0\right] $ cases are
both pathological (they do not satisfy (i) above) and will therefore be
discarded. The next simplest potential choices are $\left[ m=1,n=1\right] $
and $\left[ m=2,n=1\right] $. Both these forms, apparently, manifest no
pathologies of the previous cases. In both cases the sound speed vanishes $%
c_{s}\rightarrow0$ at vanishing density $\rho\rightarrow0$. Furthermore for $%
\rho>0$ the sound speed $c_{s}$ is initially an increasing function of $\rho 
$, as expected in reality??. One can therefore demand in each case (i.e. for
fixed $m$ and $n$ values ) that the maximum sound speed coincide with that
of light i.e. the extremum sound speed at $\frac{d^{2}p}{d\rho^{2}}=0$ be
given by $\frac{dp}{d\rho}=1$ in order to rule out superluminal behavior.
These conditions also suffice to determine the parameter $\alpha$ and allow
the fulfillment of the energy conditions in (2). Solving $\frac{d^{2}p}{%
d\rho^{2}}=0$ and $\frac{dp}{d\rho}=1$ simultaneously for the $\left[ m=1,n=1%
\right] $ case, i.e. fixes $\alpha$ at $\frac{3}{4}-\frac{1}{4}\sqrt{33}$ or 
$\frac {1}{4}\sqrt{33}+\frac{3}{4}$ (we keep only the positive solution).
Similarly for the $\left[ m=2,n=1\right] $ one fixes $\alpha$ to $\alpha
=2.\,\allowbreak213\,5$. We will select the quartic function corresponding
to $m=2,n=1$ as the representative form for our model equation of state, in
part because, unlike the cubic form, it unambiguously provides only one
possible value of $\alpha$ and makes the interpretation simpler. Thus Eq.
2.6 now takes the form 
\begin{equation}
p_{r}\left( \rho\right) =\left[ \alpha-\left( \alpha+1\right) \left( \frac{%
\rho}{\rho_{\max}}\right) ^{2}\right] \left( \frac{\rho}{\rho_{\max}}\right)
\rho,   \tag{2.7}
\end{equation}
with the constraint that $\alpha=2.\,\allowbreak213\,5$.\ It is the equation
of state we adapt throughout the remaining part of the paper.

\ \ A few words on the behavior of our assumed equation of state. While no
explicit microscopic prescription leading to its form is presently given, we
consider that its softening past the \textquotedblleft stiff" ($%
c_{s}^{2}=dp/d\rho=1$) state and its eventual conversion to an equation of
state appropriate to that of the vacuum ($p=-\rho$), is effected by the
coupling of matter to a scalar field akin to that of Higgs that gives rise
to particle masses. The dominance of the pressure by terms which render it
negative is considered to imply that eventually the energy associated with
the self interaction of this field provides the dominant contribution to the
energy momentum tensor, which is assumed to be that of a perfect fluid.

In this paper no attempt is made to follow the dynamical evolution of the
collapse. Instead it is assumed that the collapsed body has already reached
static equilibrium from its collapse. Thus, we only investigate the
characteristics of the various static 2-surfaces as the radial coordinate
decreases from the matter surface $r=R$ to the body center $r=0$. Since our
anticipated solution will allow matter fields we can, with no loss of
generality, take for initial conditions on the collapsed body surface $R$ to
be $\rho=0,\ p_{r}=0$. We therefore expect to have four regions in this
spacetime which must be satisfied by the expected solution.

Region I: Schwarzschild vacuum: $R<r<\infty$, $p=\rho=0$

Region II: Regular matter fields: $r_{\varepsilon}<r<R,\ \rho>0,0<\ p\leq
\rho,\ $

Region III: Quintessential fields: $r_{0}<r<r_{\varepsilon},\ 0<\rho
<\rho_{\max},-\rho_{\max}<p<0,\ $

Region IV: $\Lambda-$vacuum: $0\leq r\leq r_{0},\ p=-\rho=-\rho_{\max}.\ \ \ 
$

The solution to Eqs. 2.1-2.5, must satisfy the following asymptotic
conditions at large $r$ and small $r$, respectively.

(i) The spacetime must be asymptotically Schwarzschild for large $r$, i.e.
for $R<r<\infty$ \ 
\begin{equation}
ds^{2}=-\left( \ 1-\frac{2M}{r}\right) dt^{2}+\frac{1}{\left( \ 1-\frac {2M}{%
r}\right) }dr^{2}-r^{2}\left( d\theta+\sin^{2}\theta d\varphi ^{2}\right) . 
\tag{2.8}
\end{equation}
\ where for the black hole $R$\ is some hypersurface such that $R<2M$, $%
M=4\pi\int_{0}^{\infty}\rho\left( r\right) r^{2}dr\ $being the total mass.

(ii) The spacetime must be asymptotically de Sitter 
\begin{equation}
ds^{2}=-\left( \ 1-\frac{r^{2}}{r_{0}^{2}}\right) dt^{2}+\frac{1}{\left( \ 1-%
\frac{r^{2}}{r_{0}^{2}}\right) }dr^{2}+r^{2}\left( d\theta+\sin ^{2}\theta
d\varphi^{2}\right) .   \tag{2.9}
\end{equation}
for $0\leq r\leq\left( r_{0}<R\right) $. Here, $r_{0}=\sqrt{\frac{3}{8\pi
G\rho_{\max}}}=\sqrt{\frac{3}{\Lambda}}$\ signals the onset of de Sitter
behavior, where $\rho_{\max}=\rho\mid_{r\longrightarrow0}$\ is the
upper-bound on the density of the fields of order of Planck density $%
\rho_{Pl}$.

The asymptotic conditions in Eqs. 2.8 and 2.9 imply there is an interior
region which includes the family of surfaces $\Sigma=\left\{
\Sigma_{II}\cup\Sigma_{III}\right\} $ and which interfaces with region I on
the outer side and region IV on the inner side. The entire spacetime must
therefore satisfy regularity conditions at the two interfaces $r=R$\ and $%
r=r_{0}$. Put simply, such conditions guarantee (i) continuity of the mass
function and (ii) continuity of the pressure across the interfacing
hyperfaces. Thus at each interface, i.e. $\left( I,II\right) $ and $\left(
III,IV\right) $ we must have\ 
\begin{align}
\left[ \rho^{+}-\rho^{-}\right] _{\mid r=r_{i}} & =0,  \nonumber \\
&  \tag{2.10} \\
\left[ p^{+}-p^{-}\right] _{\mid r=r_{i}} & =0,  \nonumber
\end{align}
where $r_{i}=\left\{ R,r_{0}\right\} $\ and $+$, $-$\ refer to the exterior
and interior values respectively and $p=\{p_{r},p_{\perp}\}$. A desirable
feature of the model (as we find later) is the smooth continuity of the
density and pressure between region II and region III. This feature removes
the problem having to match the two regions through junction conditions
between the regular matter fields and the vacuum-like field, since here such
conditions will be satisfied naturally.

\section{The solution\ }

We now solve the Einstein equations 2.3-2.5 for the spacetime of the model.
Integration of Eq. 2.3 gives 
\begin{equation}
e^{-\lambda}=1-8\pi\frac{1}{r}\int_{0}^{r}\rho\left( r^{\prime}\right)
r^{\prime2}dr^{\prime}=1-\frac{2m\left( r\right) }{r},   \tag{3.1}
\end{equation}
where, as stated before, $m\left( r\right) $ is the mass enclosed by the a
2-sphere of radius $r$. Further, using Eqs. 2.7 and 11, integration of the $%
T_{1}^{1}$ Eq. 2.4, gives 
\begin{equation}
e^{\nu}=\left( 1-\frac{2m\left( r\right) }{r}\right) e^{\int8\pi\left[
\alpha-\left( \alpha+1\right) \left( \frac{\rho}{\rho_{\max}}\right) ^{2}%
\right] \left( \frac{\rho}{\rho_{\max}}\right) \rho\frac{dr}{1-\frac{%
2m\left( r\right) }{r}}},   \tag{3.2}
\end{equation}
where $\alpha=2.\,\allowbreak213\,5$.

\bigskip Eqs. 3.1 and 3.2 into Eq. 2.1 give 
\begin{equation}
ds^{2}=-\left( 1-\frac{2m\left( r\right) }{r}\right) e^{\Gamma\left(
r\right) }dt^{2}+\frac{1}{1-\frac{2m\left( r\right) }{r}}dr^{2}+r^{2}\left(
d\theta^{2}+\sin^{2}\theta d\varphi^{2}\right) ,   \tag{3.3}
\end{equation}
where 
\begin{equation}
\Gamma\left( r\right) =\int8\pi\left[ \alpha-\left( \alpha+1\right) \left( 
\frac{\rho}{\rho_{\max}}\right) ^{2}\right] \left( \frac{\rho}{\rho_{\max}}%
\right) \left( \frac{r}{r-2m\left( r\right) }\right) \,\rho dr   \tag{3.4}
\end{equation}
Eq. 3.3 is the general solution for the geometry of our model. It describes
the spacetime of a non-singular black hole with both matter and a de Sitter
core. We shall describe below a particular solution resulting from the
choice of a prescribed density function $\rho\left( r\right) $.

To formally complete the solution one must also solve equation 5 for the
tangential pressure $p_{\perp}=\{p_{\theta},p_{\varphi}\}$. On doing this\
one finds (see also [23] [24]) that 
\begin{equation}
p_{\perp}=p_{r}+\frac{r}{2}p_{r}^{\prime}+\frac{1}{2}\left(
p_{r}+\rho\right) \left[ \frac{Gm\left( r\right) +4\pi Gr^{3}p_{r}}{%
r-2Gm\left( r\right) }\right] ,   \tag{3.5}
\end{equation}
where $p_{r}$ is given by Eq. 2.7. Note that in this model the last term in
Eq. 3.6 is not vanishing, in general, as is the case considered in some
previous treatments where it was assumed that $p_{r}=-\rho$ for all $\rho$
(see for example [25]). Eq.3.6 is a generalization of the
Tolman-Oppenheimer-Volkoff equation [26].

One can now discuss the results of Eqs.3.1-3.4. The mass $m\left( r\right) $
enclosed a 2-surface at any radial coordinate $0<r<R$ is given by $m\left(
r\right) =4\pi\int_{0}^{r}\rho\left( r^{\prime}\right) r^{\prime
2}dr^{\prime}$. The entire mass of the body $M$ \ is given by $M=\int
_{0}^{\infty}m\left( r\right) dr=\int_{0}^{\infty}\rho\left( r\right) r^{2}dr
$. It is then clear that outside the mass of the body $T_{0}^{0}=0$.
Further, as $\rho$\ vanishes outside the body, so does the pressure (see
Eqs. 2.7 and 3.3). In particular, from Eq. 3.3 one observes that in the
limit $\rho\rightarrow0$ and $m\left( r\right) \rightarrow M$ the lapse
function and the shift vector give $e^{\nu}=e^{-\lambda}=1-\frac{2GM}{r}$.
Thus outside the mass (i.e. for large $r$) the spacetime becomes
asymptotically Schwarzschild, with the metric given by Eq. 2.8. Further, in
the limit $r\rightarrow r_{0}$, $\rho\rightarrow\rho_{\max}$ (and as both
Eqs. 2.7 and 3.4 show) $p_{r}\rightarrow-\rho_{\max}$ and $%
p_{\perp}\rightarrow-\rho_{\max }$. It follows, therefore that for $0\leq
r<r_{0}$ the fluid has constant positive density $\rho_{\max}$ and constant
negative pressure $-\rho_{\max}$ and takes on the character of a de Sitter
spacetime, through the equation of state $p=-\rho_{\max}$.

In order to establish the actual asymptotic behavior of the spacetime at
small $r$ one has to choose a density profile for the matter. We choose to
use the one employed by Dymnikova [12] because of its simplicity and
convenient form for integration over the source volume $\rho=\rho_{\max}\exp%
\left[ -\frac{r^{3}}{r_{g}\left( r_{0}\right) ^{2}}\right] $. The mass $%
m\left( r\right) $ enclosed by a 2-sphere at the radial coordinate $r$ is
then given by $m\left( r\right) =\int_{0}^{r}\rho_{\max}\{\exp\left[ -\frac{%
r^{3}}{r_{g}\left( r_{0}\right) ^{2}}\right] \}r^{\prime2}dr^{\prime}=M\left[
1-\exp\left( -\frac{r^{3}}{r_{g}\left( r_{0}\right) ^{2}}\right) \right] $,
\ where $M$ is the total mass $M=\int_{0}^{\infty}\rho_{\max}\{\exp\left[ -%
\frac{r^{3}}{r_{g}\left( r_{0}\right) ^{2}}\right] \}r^{2}dr$. The entire
mass is essentially concentrated in a region of size $%
R\simeq(r_{0}^{2}r_{g})^{1/3}$, which in general is close to $r=0$, with its
precise value depending on the value of the maximum density assumed $%
\rho_{\max}$.

The introduction of the density function $\rho=\rho_{\max}\exp\left[ -\frac{%
r^{3}}{r_{g}\left( r_{0}\right) ^{2}}\right] $ in Eq. 3.3 provides a
particular solution for the model. In the limit $r\rightarrow0$, it is seen
that the solution Eq. 3.3 asymptotically leads to Eq. 2.9 as the spacetime
becomes asymptotically de Sitter.

\section{Physical Interpretation}

Eqs. 3.3, 3.4, 3.5 along with the prescribed density profile, form the
solution in this model describing the spacetime of a non-singular black hole
containing both matter and a de Sitter core. A heuristic interpretation of
the form of the equation of state used has been given in the previous
section. The field whose presence leads to the asymptotic relation $p=-\rho$
has not been specified but it must be similar to the quintessence field used
in cosmology. In this specific case, it could in fact be the Higgs field
that gives rise to the particle masses. As such the maximum density could be
roughly $\rho_{\max }\simeq m_{H}^{4}$, the latter being the Higgs field
expectation value. This could be either of order 1 TeV or even of order of
the Plank mass if, as argued, that should be the typical mass of all scalars
coupled to gravity.

The depicted scenario allows two viable, albeit, different interpretations
in describing the internal structure of such a black hole. The two
interpretations have the common feature that in both cases a fluid with a
negative pressure, in the interior, supports the matter fields in the outer
region against any further collapse. The effect in both cases is to create a
non-singular black hole.

\subsection{The quintessential picture}

The first interpretation is directly based on the analytical functional form
of the equation of state depicted in Eq. 2.7. Here, as one moves from the
surface $R$ of the collapsed body towards the center, one first encounters
region II where from the functional form of the equation of state one
crosses a family $\Sigma_{II}$ of \ 2-surfaces with matter-like $w>0$. At
the critical point $\rho_{c}$ in the $\rho-p_{r}$ space $\frac{dp_{r}}{d\rho}%
=0$ thereafter the pressure $p_{r}$ begins to drop with increasing density $%
\rho$. At some point $\rho_{c}=\left( \frac{\alpha}{\left( \alpha+1\right) }%
\right) ^{\frac{1}{2}}\rho_{\max}$, the fluid temporarily becomes
pressureless $w\rightarrow0$, implying that $T_{1}^{1}=G_{1}^{1}\rightarrow0$%
. Beyond this, for smaller values of $r$ one enters region III in which one
is now crossing a family of 2-surfaces $\Sigma_{III}$\ for which the
equation of state is given by $-1<w<0$, \ smoothly decreasing from $w=0$
towards $w=-1$. In this picture we will refer to this as the quintessential
region. Evidently $\Sigma_{III}$ exhausts the entire family of
quintessential fluids $-1<w<0$. This fluid family provides some of the
negative pressure against the implosion of the matter fields in the outer
region. The rest of the negative pressure is provided by the constant
density $\rho_{\max}$, constant pressure $p=-\rho_{\max}$, inner core $%
0<r<r_{0}$ that mimics the cosmological constant. This interpretation is
valid provided the equation of state $p_{r}\left( \rho\right) $ is, indeed
in reality, a well-behaved function of $r$ for $\rho_{c}<\rho<\rho_{\max}$.

In this quintessential picture, one can estimate the amount remaining pure
matter (the total field for which $w\geq0$) as a fraction of the total
energy of the system. This matter lies in the outer region $%
r_{\varepsilon}\leq r\leq R$ (i.e. between points B and C in Figure 1).
Since B is given by, $\frac{dp_{r}}{d\rho}\mid_{\rho_{c}}=0$ and C is given
by $p_{r}\mid _{\rho_{\varepsilon}}=0$, one can infer from Eq. 7 that in
this region the density function $\rho$ is bounded by $\left[
\rho_{c}=\left( \frac{\alpha }{3\left( \alpha+1\right) }\right) ^{\frac{1}{2}%
}\rho_{\max}\right] \leq\rho\leq\left[ \rho_{\varepsilon}=\left( \frac{\alpha%
}{\left( \alpha+1\right) }\right) ^{\frac{1}{2}}\rho_{\max}\right] $. To
determine the matter mass $_{\mathcal{M}matter}$ enclosed we use the density
profile $\rho\left( r\right) =\rho_{\max}\exp\left[ -\frac{r^{3}}{%
r_{0}^{2}r_{g}}\right] $ and perform the integral $\int\rho r^{2}dr$ with
the appropriate limits. It is useful to change the independent variable from 
$r$ to $\rho$. Then $r^{2}dr=\frac{1}{3}r_{0}^{2}r_{g}\left( \frac{\rho_{m}}{%
\rho}\right) d\left( \frac{\rho}{\rho_{m}}\right) $ and we have that $_{%
\mathcal{M}matter}=4\pi\int\rho r^{2}dr=\frac{4}{3}\pi
r_{0}^{2}r_{g}\int_{0}^{\rho_{\varepsilon}}d\rho$. Thus 
\begin{equation}
\mathcal{M}_{matter}=\frac{4}{3}\pi r_{0}^{2}r_{g}\left( \frac{\alpha}{%
\alpha+1}\right) ^{\frac{1}{2}}\rho_{\max}.   \tag{4.1}
\end{equation}

Using $\alpha=2.\,\allowbreak213\,5$ as adopted earlier for our model, gives 
$M_{matter}=\allowbreak0.829\,95\left( \frac{4}{3}\pi r_{0}^{2}r_{g}\right)
\rho_{\max}$. On the other hand the total mass $M=4\pi\int\rho\exp\left[ 
\frac{r^{3}}{r_{0}^{2}r_{g}}\right] r^{2}dr=\frac{4}{3}\pi
r_{0}^{2}r_{g}\rho_{\max}$. Thus the fractional content of matter in the
quintessential picture for the black hole is $\frac{\mathcal{M}_{matter}}{M}%
=\allowbreak 0.829\,95\simeq0.83$.

\subsection{A two-fluid system}

One can attempt yet a seemingly equally viable interpretation of the black
hole's internal structure. If one holds as in [6] the view of a phase
transition of the fluid from the matter-like form to the de Sitter state
given by $p=-\rho_{\max}$, then the successive 2-surfaces in the family $%
r_{0}<r<r_{c}$ will represent a two fluid system of matter fields and the $%
\Lambda-$vacuum which in both dynamic and thermodynamic equilibrium. The
fluid begins as pure matter on the 2-surface $r_{c}$ and gets richer and
richer in the $\Lambda-$vacuum state as one moves deeper and deeper into the
black hole. Then the density $\rho_{r}\left( r\right) $ at any such point on
any such 2-surface shell $r_{0}<r<R$ is really a sum of two partial
densities 
\begin{equation}
\rho=\beta\left( r\right) \rho_{c}-\left[ 1-\beta\left( r\right) \right]
\rho_{\max},   \tag{4.2}
\end{equation}
where $\rho_{c}$ is the critical density just before collapse when $p=\rho
_{c}$ and $0\leq\beta\leq1$. The function $\beta\left( r\right) $ is the
fractional content of pure matter fields to the total surface energy
contained in an elementary 2-surface shell $4\pi r^{2}dr$ at $r_{0}<r<R$.
For decreasing $r$, $\beta\left( r\right) $ is a decreasing function,
evolving from $\beta\left( R\right) =1$ to $\beta\left( r_{0}\right) =0$. In
terms of the partial pressures of the matter fields and the de Sitter
vacuum, the pressure at any position $r_{0}\leq r<R$ is then given by 
\begin{equation}
p_{r}\left( r\right) =\underset{n=0}{\overset{3}{\sum}}a_{n}\left( \frac{%
\rho_{c}}{\rho_{\max}}\right) ^{n}\left[ \beta\left( r\right)
\rho_{c}-\left( 1-\beta\left( r\right) \right) \rho_{\max}\right]   \tag{4.2}
\end{equation}
where the $a$'s are positive constants given respectively by 
\begin{align}
a_{0} & =\left( 1-\beta\right) \left[ \alpha-\left( \alpha+1\right) \left(
1-\beta\right) ^{2}\right] ,  \nonumber \\
a_{1} & =\beta\left[ \alpha+\left( \alpha+1\right) \left( 1-\beta \right) %
\right] ,  \nonumber \\
a_{2} & =\beta^{2}\left( \alpha+1\right) \left( 1-\beta\right) ,  \tag{4.4}
\\
a_{3} & =\beta^{3}\left( \alpha+1\right) .  \nonumber
\end{align}
This expression based on partial pressures will then replace Eq. 2.7 as the
equation of state $p_{r}\left( \rho\right) $ in the event the evolution of
matter fields to a de Sitter state involves a spontaneous phase transition.
One notes that the pressure $p_{r}$ still has the desired asymptotic
behavior. Thus in the limit $\beta\left( r\right) \rightarrow1$, $%
p\rightarrow p_{c}=\left[ \alpha-\left( \alpha+1\right) \left( \frac{\rho_{c}%
}{\rho_{\max}}\right) ^{2}\right] \left( \frac{\rho_{c}}{\rho_{\max}}\right)
\rho_{c}$, while in the limit $\beta\left( r\right) \rightarrow0$, $%
p\rightarrow-\rho_{\max}$ as expected.

One can now estimate the energy $M_{matter}$ in the pure matter fields as a
fraction of the total energy of the black hole in this two-fluid picture. We
assume that up to the critical density $\rho_{c}$ the matter is pure and
that the energy content contribution of the de Sitter fluid with an energy
density $\rho_{\max}$ starts to grow at $\rho=\rho_{c}$. Then the matter
content can be written as a two piece integral $M_{matter}=4\pi%
\int_{r_{c}}^{\infty}\rho r^{2}dr+4\pi\int_{r_{0}}^{r_{c}}\beta\left(
r\right) \rho_{c}r^{2}dr$. We will assume $\beta\left( r\right) $ to be
simply linear in $r$ with a form $\beta\left( r\right) =\frac{r-r_{0}}{%
r_{c}-r_{0}}$, which satisfies the requirements imposed on it above. Then
substituting for $\beta\left( r\right) $ and again noting that $r^{2}dr=%
\frac{1}{3}r_{0}^{2}r_{g}\left( \frac{\rho_{m}}{\rho}\right) d\left( \frac{%
\rho}{\rho_{m}}\right) $ we get 
\begin{equation}
\mathcal{M}_{matter}=\frac{4}{3}\pi\left\{ \left[ r_{0}^{2}r_{g}\right] +%
\left[ \frac{1}{4}\left( r_{c}+r_{0}\right) \left(
r_{c}^{2}+r_{0}^{2}\right) -\frac{1}{3}r_{0}\frac{r_{c}^{3}-r_{0}^{3}}{%
r_{c}-r_{0}}\right] \right\} \rho_{c}.   \tag{4.5}
\end{equation}

From the definition of the density function we have that $r=\left[
-r_{0}^{2}r_{g}\ln\left( \frac{\rho}{\rho\max}\right) \right] ^{\frac{1}{3}}$%
. Since $r\rightarrow r_{0}$ as $\rho\rightarrow\rho_{\max}$ we have that
for $0<r<r_{0}$, $r\left( \rho\right) $ is not well defined since $r$ is now
not single-valued in $\rho$. Further, we don't know the actual size of $r_{0}
$. As a result we shall assume, for the purpose of evaluating Eq. 4.5, that
both $r=\left[ -r_{0}^{2}r_{g}\ln\left( \frac{\rho}{\rho\max}\right) \right]
^{\frac{1}{3}}$ and $r\rightarrow r_{0}$ as $\rho\rightarrow\rho_{\max}$
hold so that $r_{0}\left( \rho_{\max}\right) \rightarrow0$. This is
equivalent to having the de Sitter vacuum approached only as $r\rightarrow0$%
. Thus the value of $M_{matter}$ calculated from here will be an upper
bound. Applying this on the integral $\int_{r_{0}}^{r_{c}}\beta\left(
r\right) \rho_{c}r^{2}dr$ part Eq. 4.5 implies $\left[ \frac{1}{4}\left(
r_{c}+r_{0}\right) \left( r_{c}^{2}+r_{0}^{2}\right) -\frac{1}{3}r_{0}\left( 
\frac{r_{c}^{3}-r_{0}^{3}}{r_{c}-r_{0}}\right) \right] \rightarrow\frac{1}{4}%
r_{c}^{3}\rho_{c}$ so that Eq. 4.5 reduces to 
\begin{equation}
\mathcal{M}_{matter}=\frac{4}{3}\pi\left[ r_{0}^{2}r_{g}+\frac{1}{4}r_{c}^{3}%
\right] \left( \frac{\alpha}{3\left( \alpha+1\right) }\right) ^{\frac{1}{2}%
}\rho_{\max}   \tag{4.6}
\end{equation}
Use of $r_{c}=\left[ -r_{0}^{2}r_{g}\ln\left( \frac{\rho_{\varepsilon}}{%
\rho\max}\right) \right] ^{\frac{1}{3}}$ and $\alpha=2.\,\allowbreak213\,5$
then gives $\frac{\mathcal{M}_{2f}}{M}=0.479\,17\left[ 1+\allowbreak
0.735\,70\right] =0.8316\simeq0.83$.

This result agrees with the one found above in the quintessential picture.

\section{Conclusion}

In this paper we have suggested a possible model for a non-singular black
hole as a product of gravitational collapse. It is an exact solution of the
Einstein Equations, being Type [II, (II)] by Petrov classification. This
model is based on a judicious choice we have made of the equation of state
of the collapsed matter. At high densities this equation of state violates
some of the energy conditions [2][3] originally used to justify the
existence of black hole singularities. In our model this violation leads to
non-singular collapse. On the other hand, for the entire density parameter
space the equation of state in our approach still satisfies the Weak Energy
Condition $\rho\geq0$, $\rho+p_{r}\geq0$, a basic requirement for physical
fields.

The solution depicts a spacetime with matter fluids in the outer layers
(region II) which, as one moves deeper inside, give way to either a
quintessential or a two-fluid region III. Region III, in turn, evolves into
region IV, an inner-most core with de Sitter characteristics. The fluids in
regions III and IV both provide the negative pressures needed to sustain the
outer matter in static equilibrium. The solution has the required asymptotic
forms, reducing to the Schwarzschild vacuum solution outside the matter
fields and reducing to the de Sitter solution as one approaches the black
hole center. The existence of region III renders the interface between
matter fields and the de Sitter vacuum to join smoothly. This is because
both the density $\rho$, the radial pressure $p_{r}$ and the tangential
pressure $p_{\perp}$ profiles and their derivatives are continuous. This
character of the fields makes the matching across associated interfaces
natural.

Using our model we have, in Section IV, offered two viable interpretations
about the possible macro-state of the fields making up the total energy of
the black hole. These interpretations are based on two pictures, namely a
Quintessential Picture and a Two Fluid Picture. In both cases we have
estimated the fractional contribution of the regular matter-like fields as a
fraction of the total black hole energy. We find for an upper bound\ the
same value of $\sim0.83\ $in both cases\footnote{%
It is remarkable that the two different interpretations yield the same
numerical values of the mass-fractions.}. As a corollary, our model suggests
that the lower bound for the amount of matter that is found in the
\textquotedblleft exotic" state can be as small as $17\%$ of the entire
configuration of the collapsed matter. Both the size and density of this de
Sitter-like central core region will depend on the details of the
microscopic physics that lead to the specific equation of state we have
employed, which go beyond the scope of the present work. However, these
considerations are consistent with our present notions of dynamical particle
masses and symmetry breaking, as the density of this state is much higher
than those produced to-date in the laboratory. The precise value is
uncertain but it has to be at least as high as $m_{H}^{4}$, where $m_{H}$ is
the mass of the Higgs field, and it can be as high as the Planck density.
Our entire non-singular configuration is well within the horizon of the
black hole and in this respect the configuration is very different from some
previous treatments, e.g. [6]. In these treatments, the field configuration
fills the entire volume interior to $r=2M$ with a fluid with $p=-\rho$,
whose energy density for sufficiently large black holes can be smaller than
that of water. This raises the difficult problem of converting matter from
the usual equations of state to the \textquotedblleft exotic" $p=-\rho,$
under usual laboratory conditions.

Finally, the model leaves some issues unresolved. First, the calculation for
these mass fractions is based on assuming that in the region $r_{0}<r<R$
both the pressure and the density are well-behaved, functions in the radial
coordinate $r$, so that conversely $r\left( \rho\right)
=-r_{0}^{2}r_{g}\ln\left( \frac{\rho}{\rho_{\max}}\right) $ well defined in
the entire parameter space of $\rho$. Since in our model $%
\rho\rightarrow\rho_{\max}$ when $r\rightarrow r_{0}$, then $r\left(
\rho\right) $ vanishes at $r_{0} $ forcing (in the mass calculation) the de
Sitter state to appear virtually at the origin. It is in this sense that $%
\frac{\mathcal{M}_{matter}}{M}=0.83$ is an upper-bound. One can reduce this
fraction by using $r\left( \rho\right) =-r_{0}^{2}r_{g}\ln\left( \frac{\rho}{%
\rho_{\max}}\right) +r_{0}$ in the calculations, instead. However since we
don't know the radial extent the de Sitter field could fill at the core, or
whether indeed it fills any space bigger than the Planck length we can not
presently constrain $\frac {\mathcal{M}_{matter}}{M}\ $from below. This is
one issue for quantum gravity to settle.

Secondly, our classical model can not distinguish between the two pictures
offered to determine which choice of the matter macrostate is the correct
one. However, the fact that in the region $r_{0}<r<R$ the energy-momentum
tensor elements $\rho$ and $p$ are well-behaved functions of the radial
coordinate $r$ may provide an interesting insight. It suggests that one can
associate a unique value of the energy-momentum tensor on each 2-surface in
the family $\Sigma=\left\{ \Sigma_{II}\cup\Sigma_{III}\right\} $. One can
further speculate that, such a configuration may even be quantizable. In
such a case the information pertaining to the pre-collapse phase of the
object would not be lost but would now reside on the (quantized) onion-like
family of hyperfaces $\Sigma$ inside the black hole. This is another issue
for quantum gravity to settle. These results suggest a need for further
investigation.

\ \ 

\begin{acknowledgement}
This research was performed while one of the authors (MRM) held a National
Research Council Senior Research Associateship Award at (NASA-Goddard).
\end{acknowledgement}

\newpage

\begin{figure*}[hbt]
\centerline{\epsfig{file=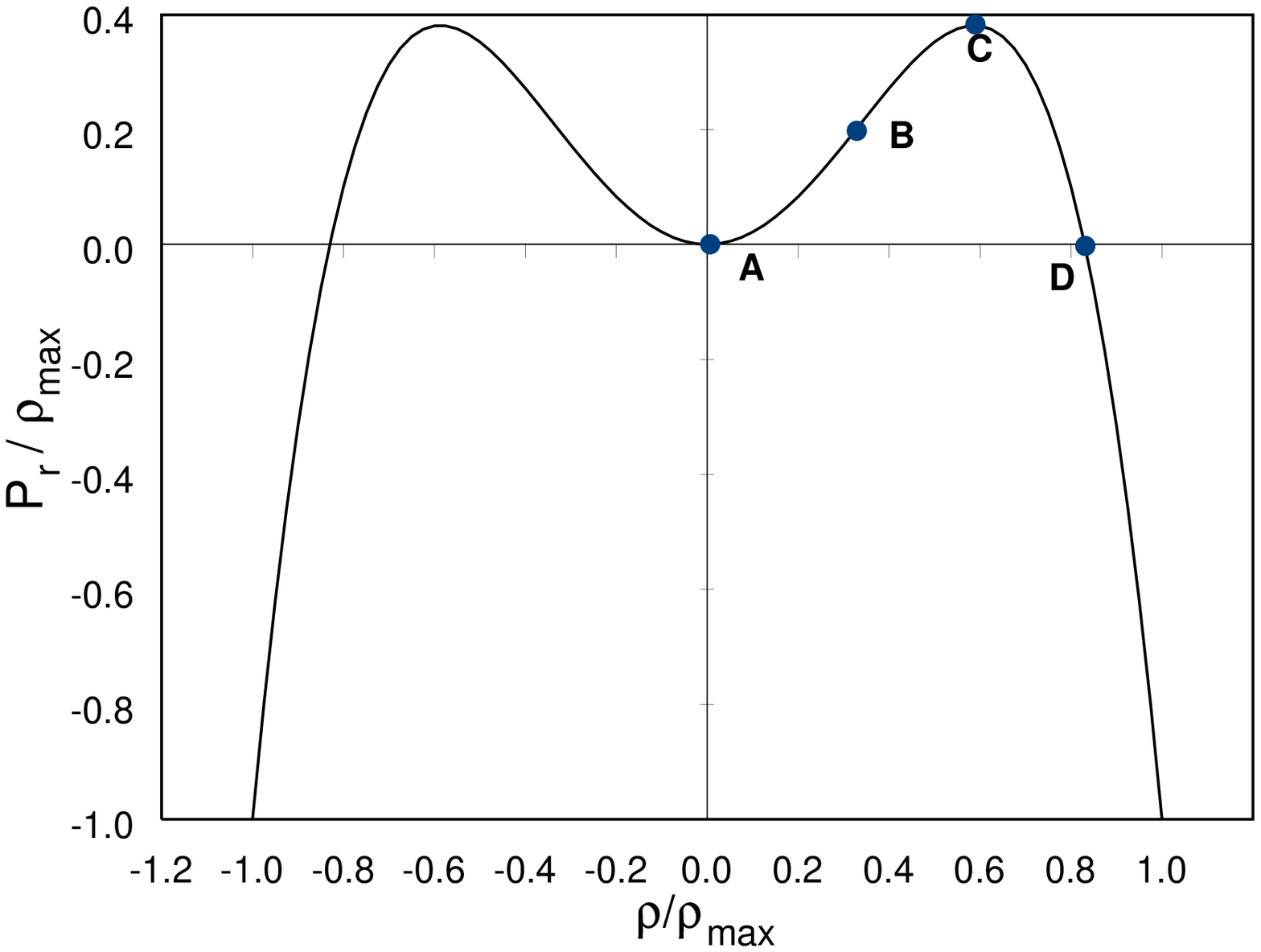,height=10.0cm,width=10cm}}  
\caption{\small \ The quartic equation of state functional }$%
p\left( \rho \right) ${\small \ with }$m=2,~n=1${\small .}
{\small At A when }$\rho =0${\small , then }$p=0${\small \ and }$\frac{dp}{%
d\rho }=0${\small . The value of the parameter }$\alpha =2.213\left(
5\right) ${\small \ is chosen so that for the entire density parameter space 
}$\rho \geq 0${\small \ the sound speed is not superluminal. The point B
giving the maximum slope corresponds to the maximum sound speed, }$\frac{%
d^{2}p}{dr^{2}}=0${\small , chosen to correspond to the light speed, }$\frac{%
dp}{d\rho }=c^{2}=1${\small . Point C is the critical point }$\left( \rho
_{c},p_{c}\right) ${\small , and for }$\rho >\rho _{c}${\small , then }$%
p\left( \rho \right) ${\small \ is a decreasing function. At point D or }$%
\rho =\rho _{\varepsilon }${\small , the pressure temporarily vanishes..}
\end{figure*}

\end{document}